# The Tidal Circularization Cutoff Period
# of the Old Open Cluster NGC 188[1]

Robert D. Mathieu, Søren Meibom, and Christopher J. Dolan

Department of Astronomy, University of Wisconsin – Madison, Madison WI 53706
mathieu@astro.wisc.edu, meibom@astro.wisc.edu, dolan@clotho.com

Running Title: NGC 188 Tidal Circularization Cutoff Period

---

[1] WIYN Open Cluster Study. XVIII.

## Abstract

Based on a sample of 45 main-sequence solar-type spectroscopic binaries we find the tidal circularization cutoff in the open cluster NGC 188 (age 7 Gyr) to be well-defined at a period of 15.0 days. When compared to shorter tidal circularization cutoff periods in the younger clusters M67 and the Hyades, this result establishes that tidal circularization processes are effective in solar-type main-sequence binaries for timescales longer than 1 Gyr. This result is in contrast to the theoretical prediction by Zahn & Bouchet that tidal circularization due to equilibrium tides is ineffective for late-type main sequence stars, and that tidal circularization cutoff periods are instead set only by pre-main-sequence tidal circularization. The NGC 188 cutoff period is also longer than can be explained by present theories of tidal circularization due to dynamical tides.

Subject headings: binaries : close --- binaries: spectroscopic --- stars : evolution

# 1. Introduction

Tidal interactions between stars play central roles in the evolutionary stories of close binary systems, and as such have been important in our understanding of a broad range of stellar phenomena, including rotation, binary orbits, abundance evolution, blue stragglers, cataclysmic variables, X-ray binaries, and Type I supernovae. The physics of stellar tidal interactions has been thoroughly developed, for example by Zahn and Hut (e.g., Zahn 1992, Hut 1981). However, in the context of equilibrium tide models the timescales of rotational synchronization and orbital circularization rest on still poorly understood dissipation mechanisms. For late-type stars this dissipation is often attributed to turbulent friction in convective zones (see for example Zahn 1977, Zahn 1989, and Goldman & Mazeh 1991). Alternatively, circularization through dynamical tides and resonant interactions in the radiative cores of late-type stars has been predicted to be more effective than equilibrium tides (Savonije & Witte 2002; Witte & Savonije 2002; Goodman & Dickson 1998). Finally, Tassoul (e.g., Tassoul 2000) has proposed hydrodynamical flows resulting from non-synchronous tidal distortions as the path to synchronization and circularization.

Within the context of the equilibrium tide model, Zahn (1989) concluded that convective turnover timescales were too long to permit significant dissipation during the periastron passages of short-period binaries. Building on this conclusion, Zahn & Bouchet (1989, hereinafter ZB) made the provocative assertion that for late-type binaries essentially all of the significant tidal interaction prior to the giant phase of evolution occurs during pre-main-sequence evolution when such stars have large radii and deep convective zones. Despite the short duration of pre-main-sequence evolution, the dramatic increase of tidal torques with large stellar radius permits effective synchronization and circularization. Equally importantly, ZB found that dissipation mechanisms during the main-sequence phase of evolution are inadequate to drive significant synchronization or circularization despite the Gyr timescales available.

A critical observational prediction of ZB was that all late-type main-sequence binaries should have circular orbits for periods less $\approx$ 8 days and (primordial) eccentric orbits at longer periods, independent of age. ZB predicted some variation in cutoff period as a function of the stellar masses, but the range of variation ($\approx$1 day) was small, and substantially less than previously predicted variations of cutoff periods with stellar age (Mathieu & Mazeh 1988).

Contemporaneous with this theoretical work, observers were finding in late-type main-sequence binary populations well-defined transitions between circular orbits at short periods and eccentric orbits at longer periods. The transition periods were defined as "tidal circularization cutoff periods" (Mayor & Mermilliod 1984, Mathieu & Mazeh 1988, Duquennoy *et al*. 1992). Cutoff periods found among main-sequence binaries in the Hyades (0.6 Gyr; 5.6-8.5 days), M67 (5 Gyr; 12.4 days) and the galactic halo (10 Gyr; 20 days) increased with age, seeming to indicate ongoing tidal circularization on the main sequence. On the other hand, a poorly constrained cutoff period of 7.0 days in the Pleiades (0.1 Gyr) suggested little orbital evolution among binaries with ages less than 1 Gyr, and more recently Melo et al. (2001) have argued that the pre-main-sequence tidal circularization cutoff period is also roughly 7 days.

Mathieu *et al*. (1992) provided a comprehensive discussion of the interface between theory and observations, and suggested that the data were consistent with a hybrid model in which effective tidal circularization during the pre-main-sequence phase sets an "initial" cutoff period at $\approx 8$ days which remains unchanged for roughly 1 Gyr. After the passage of 1 Gyr the action of main-sequence tidal processes are able to circularize the orbits of binaries with periods longer than 8 days, leading to an increase in tidal circularization cutoff periods with binary population age.

However this qualitative conjecture had no physical mechanism that explained both branches of the hybrid model. The dissipation model of Zahn (1989) predicted the observed cutoff periods for young binary populations, but could not at the same time explain ongoing main-sequence tidal circularization beyond 1 Gyr. They calculations of Savonije & Witte showed dynamical tides to be more effective, but only in cases of unrealistically slow stellar rotation could they explain the observed cutoff periods beyond 1 Gyr. Finally the models of Tassoul can match the observations, but only through proper calibration of free parameters.

Determinations of tidal cutoff periods among old binary populations are critical for setting the timescale of main-sequence tidal circularization . Given the possibly different pre-main-sequence evolutionary history of the metal-poor halo stars (ZB), the observational evidence for ongoing main-sequence tidal circularization has rested largely on a single old cluster, M67. This Letter presents a newly derived tidal cutoff in another old open cluster, NGC 188. The well-determined tidal cutoff of 15.0 days in a cluster of age 7 Gyr supports ongoing main-sequence tidal circularization on timescales in excess of 1 Gyr.

## 2. Observations

NGC 188 is among the richest of old open clusters in the Galaxy. Recent studies have derived an age of 7 Gyr (Sarajedini et al. 1999) with a metal abundance very near solar (e.g., Twarog, Ashman & Anthony-Twarog 1997). As such, the binaries of NGC 188 are an excellent sample for a comparison of orbital properties with previously studied late-type binary populations.

In 1995 we began a high-precision radial-velocity survey of NGC 188 using the WIYN Observatory[2] Multi-Object Spectrograph (Mathieu *et al.* in preparation). Typical single-measurement precisions are 0.5 km s$^{-1}$ (Meibom et al. 2001). The stellar sample was selected from the proper-motion survey of Dinescu *et al.* (1996). We have observed 512 stars selected as proper-motion members within the magnitude range 12.0 < V < 16.5. This range includes most of the giant branch, blue stragglers, the subgiant branch, and the uppermost 1.5 magnitudes of the cluster main sequence. The last are solar-type stars, and are the basis for this Letter.

---

[2] The WIYN Observatory is a joint facility of the University of Wisconsin – Madison, Indiana University, Yale University, and the National Optical Astronomy Observatories.

All stars have been observed at least 3 times, with most observed more often. Binary systems were monitored as soon as identified, and to date orbital solutions have been obtained for 45 main-sequence solar-type spectroscopic binaries. The complete solutions will be presented and discussed in Mathieu *et al.* (in preparation); here only the orbital periods and eccentricities are needed.

### 3. Period-Eccentricity Distribution

In Figure 1 we show the distribution of orbital eccentricity versus orbital period for solar-type binaries in NGC 188 with periods less than 50 days. For longer periods the orbital eccentricities are distributed over a wide range from 0.1 – 0.7, with a notable absence of circular orbits. In contrast, with one exception all binaries with periods less than 15 days have orbital eccentricities consistent with zero. (For simplicity, hereinafter such binaries will be said to have circular orbits.) There are four circular binaries – with Dinescu *et al.* identifications of D55, D394,

D1255, D1376 - with orbital periods between 14 days and 15 days, assuring that the cutoff period is not defined by an aberrant case. We follow Duquennoy *et al.* (1992) and define the tidal circularization cutoff period $T_{cutoff}$ as the period of the longest period circular orbit. Thus the tidal circularization cutoff period of NGC 188 is $T_{cutoff}$ = 15.0 days, this being the orbital period of D1376 (P = 14.991 ± 0.001 days, e = 0.018 ± 0.014).

## 4. Discussion

In Figure 2 we show the now standard figure of log $P_{cutoff}$ vs log T, where T is the age of each coeval stellar population. Clearly, the NGC 188 tidal circularization cutoff period is substantially larger than cutoff periods found in the stellar populations with ages of less than 1 Gyr. The cutoff period of the Hyades is somewhat ambiguous; among solar-type binaries the cutoff period is 5.6 days, but there is a 0.5 $M_o$ star with a circular orbit at 8.5 days. Both cutoff periods are indicated in Figure 2, and the conclusion is the same whichever is adopted here. This observation straightforwardly supports the conclusion of Mathieu *et al.* (1992) that main-sequence tidal circularization does proceed on timescales in excess of 1 Gyr.

The defining binary – D394 – is located on the unevolved main sequence (V = 15.7, B-V = 0.70; Sarajedini *et al.* 1999). Given the sensitivity of tidal circularization to stellar radius, this property of D394 is important in assuring a secure measure of the cutoff period. The other three binaries are somewhat brighter . Their primaries may be slightly evolved, although the double-lined nature of S55 and the location of S1255 above the main sequence argue that their composite luminosities are overestimates for the primary light.

The star D1395 is a puzzle, having an orbital eccentricity of e=0.35 at a period of 10.2 days. As with all of the binaries discussed here, D1395 is a proper-motion, radial-velocity, and photometric member. D1395 is somewhat fainter than D394 and thus the primary is also unevolved. Duquennoy *et al.* (1992) suggested that such stars had initially large eccentricities and are still in the process of being circularized.

The NGC 188 cutoff period is consistent with the slope previously indicated by the Hyades, M67 and halo cutoff periods. The data are too few to put much weight on this trend, but we note that the slope is steeper than $\gamma = 16/3$, as would be derived in the equilibrium tide scenario from purely dynamical considerations (see Mathieu *et al.* 1992 for an overview). The steeper slope is consistent with a decrease in dissipation efficiency with shorter periods, as suggested by Zahn (1989; see also Goldman & Mazeh 1991) due to convective turnover timescales becoming commensurate with the orbital periods. The hydrodynamical mechanism of Tassoul also suggests a steeper slope.

## 5. Summary

We have determined orbital solutions for 45 solar-type main-sequence binaries in the old open cluster NGC 188. The short period binaries clearly define a tidal circularization cutoff at 15.0 days. Thus NGC 188 joins a progression of increasing cutoff periods with increasing age among late-type main-sequence binaries in the Hyades (5.6-8.5 days, 0.6 Gyr), M67 (12.4 days, 5 Gyr), NGC 188 (15.0 days, 7 Gyr) and the halo (20 days, 10 Gyr).

We conclude from this progression that tidal circularization processes are effective in late-type main-sequence binaries with periods in excess of 8 days on timescales longer than 1 Gyr. This result is in contrast to the ZB prediction that tidal circularization only occurs during the pre-main-sequence phase, and continues to suggest the hybrid scenario put forth by Mathieu et al. (1992; see recent review by Melo *et al.* 2001) in which the circular orbits in young binary populations are the result of pre-main-sequence tidal circularization while the longer period circular orbits in older populations are due to main-sequence tidal circularization. However, the Pleiades cutoff period is poorly determined due to the small sample size; we remain very much in need of well determined cutoff periods for young clusters. More fundamentally, as yet no physical mechanism of tidal circularization has reproduced the distribution presented in Figure 2.

We would like to thank Nick Stroud for his work in deriving many of the orbital solutions on which this paper is based, and to express our appreciation for the superb long-term support of the WIYN Observatory staff. This work has been supported by NSF grant AST 97-31302 and by a Ph.D. Fellowship from the Danish Research Agency to SM.

Figure 1   Orbital eccentricity vs. period for short-period main-sequence solar-type binaries in NGC 188. Error bars are 1 σ; uncertainties in periods are smaller than the symbols. Note the distinct transition from circular to eccentric orbits defining a tidal circularization cutoff period at P=15.0 days.

Figure 2   Tidal circularization cutoff period vs. age for six coeval stellar samples: pre-main-sequence ( Melo *et al*. 2001); Pleiades (Mermilliod *et al*. 1992); Hyades (Duquennoy *et al*. 1992; see text for explanation of two values); M67 (Latham *et al*. 1992); NGC 188; Milky Way halo stars (Latham *et al*. 2002). Note the evident increase in tidal circularization period with age beyond 1 Gyr, indicating ongoing tidal circularization for main-sequence solar-type binaries.

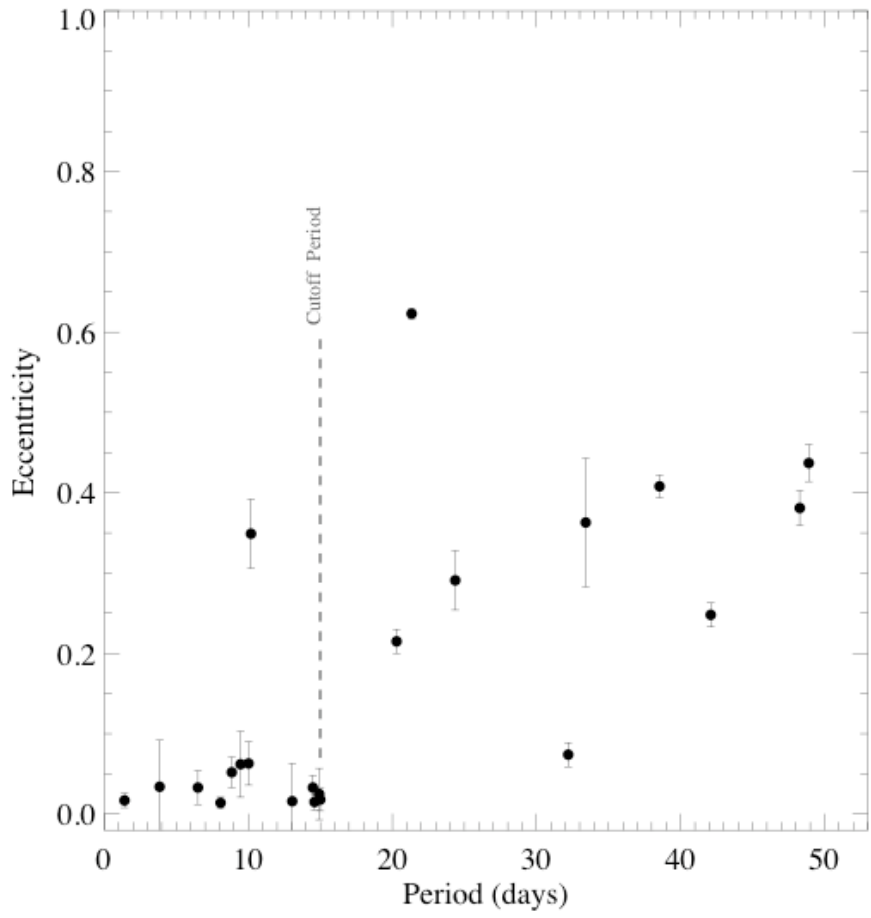

Figure 1

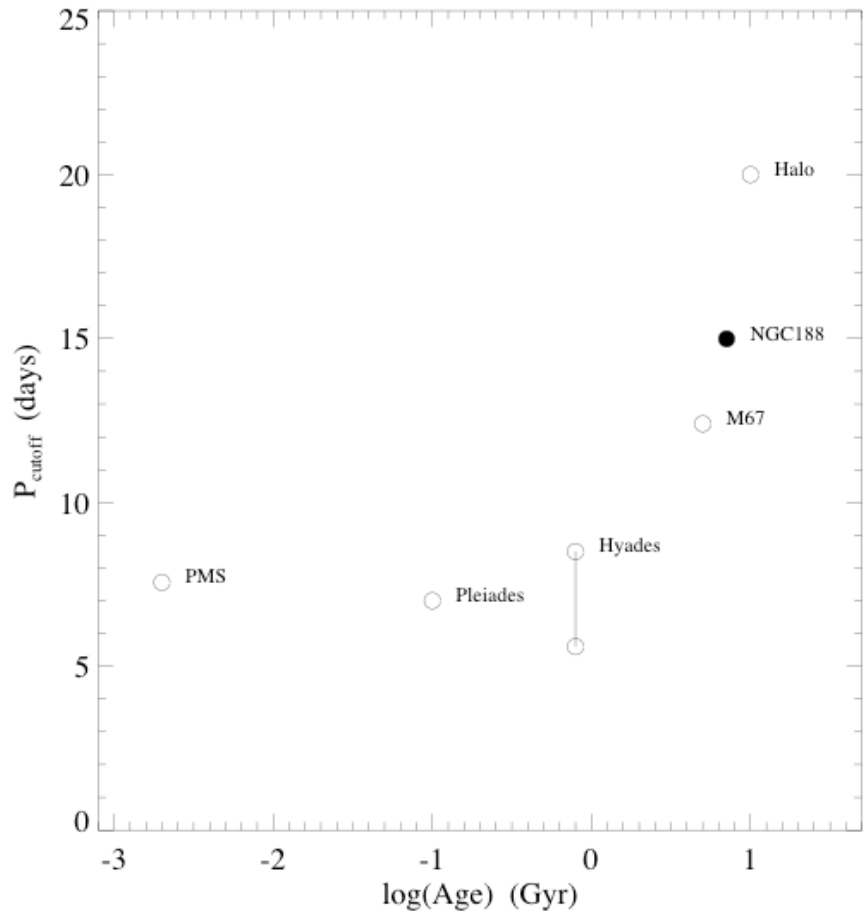

Figure 2